\documentclass[aps,prl,showpacs,twocolumn]{revtex4}
\usepackage{amssymb}
\usepackage{graphicx}
\usepackage{color}

\begin{document}

\title{Diffusion of particles in simple fluids: A joint theory of kinetics and hydrodynamics}
\author{Hanqing Zhao$^{1,2}$}
\author{Hong Zhao$^{1,3}$}
\email{zhaoh@xmu.edu.cn}
\affiliation{$^{1}$Department of Physics and Institute of Theoretical Physics and
Astrophysics, Xiamen University, Xiamen 361005, Fujian, China\\
$^{2}$Department of Modern Physics, University of Science and Technology of
China, Hefei 230026, Anhui, China\\
$^{3}$Collaborative Innovation Center of Chemistry for Energy Materials,
Xiamen University, Xiamen 361005, Fujian, China}

\begin{abstract}
The particle diffusion in a fluid is a classical topic that dates back to
more than one century ago. However, a full solution to this issue still
lacks. In this work the velocity autocorrelation function and the diffusion
constant are derived analytically, and the hydrodynamics effect on the
particle diffusion is analyzed in detail. Unlike previous studies, the
ring-collision effect is exhaustively considered in our treatment, and the
hydrodynamics approach is extended to the whole time range. Large scale
molecular dynamics simulations for the hard-disk fluid show that our
analytical results are valid up to the density close to the crystallization
point.
\end{abstract}

\date{\today }
\maketitle

The particle diffusion in a fluid is a profound issue of statistics physics.
A primary attack to this problem was made by Boltzmann in 1872, and a more
comprehensive understanding was obtained due to Einstein's study~\cite
{Einstein} in 1905 to the Brownian motion. These seminal studies
established the kinetics theory, which predicts that the velocity
autocorrelation function (VACF) of a particle decays exponentially, implying
immediately a time-independent diffusion constant~\cite{theorySP, Leener}.
However, it was found later by Alder and Wainwright in 1967 that the VACF may
feature a long-time tail~\cite{Alder}, suggesting that the kinetics theory
is not thorough yet. Since then the long-time tail has been intensively
studied~\cite{theorySP, Leener, Alder, Alder2, Alder3, wain-tlnt, Dorfman1,
Dorfman2, Dorfman3, Ernst, tlnt, cv-longtail, review-l-t-coupling,
high-density, formula-2d, 2d-tailnotimport, Erpenbeck} and has been
attributed to the hydrodynamics effect. To harmonize these results, the
diffusion process is divided into two stages, the kinetics stage (KS) for
short time and the hydrodynamics stages (HS) for long time~\cite{theorySP},
and the VACF reads
\begin{equation}
{C(t)}/{\Omega }=\left\{
\begin{array}{ll}
\exp ({-\Omega t/D_{k}}), & \text{KS;} \\
\sim \lbrack (1-1/d)/n][4\pi (D_{k}+\nu _{k})t]^{-\frac{d}{2}}, & \text{HS}.%
\end{array}%
\right.  \label{1}
\end{equation}%
Here $t$, $n$, $D_{k}$, $\nu _{k}$, and $d$ are, respectively, the
correlation time, average particle number density, kinetics diffusion
constant, kinetics viscosity diffusivity, and dimensionality of the system; $%
{\Omega =C(0)=k_{B}T/m}$ ($k_{B}$ and $T$ are the Boltzmann constant and the
system temperature; $m$ is the mass of particles). The hydrodynamics
consideration provides a correction to Einstein's picture, and as a result,
the diffusion constant, denoted by $D(t)$, turns out to be time dependent
following the Green-Kubo formula
~\cite{theorySP, Kubo, Andrieux}; In particular, the correction to the
diffusion constant is finite for $d=3$ but makes the diffusion constant
divergent in the thermodynamical limit for $d=2$.

Despite of these developments, how a particle diffuses in a fluid is still
unclear. On one hand, qualitatively, the tail of the VACF may not follow the
predicted power law. For example, for $d=2$, it has been shown that the
asymptotic self-consistent solution of the VACF is $C(t)\sim (t\sqrt{\ln t})^{-1}$
instead~\cite{wain-tlnt, tlnt}. In a recent careful numerical study of the
hard-disk fluid, it has been found that the VACF tends to $\sim t^{-1}$ at
low densities but $\sim (t\sqrt{\ln t})^{-1}$ at moderate
densities~\cite{pre2008}. On the other hand, quantitatively, the amplitude
of the power-law tail given by the hydrodynamics theory suffers from a
cut-off approximation in the wave-vector space~\cite{theorySP}; in addition,
the transition time from the KS to the HS has not been identified precisely
either. All these limitations make it impossible to accurately evaluate the
hydrodynamics effect based on the existing theory.

In clear contrast, the experimental techniques have been developed so fast
in recent years that nowadays it has been possible to measure the
instantaneous velocity of a Brownian particle in laboratories~\cite{Science,
Nature, WangBo}. More importantly, the hints of the hydrodynamics effect
have been evidenced \cite{Nature}. It is thus highly desired to give the
particle diffusion process a thorough theroy. Motivated by this, in the
present work we derive the VACF and the diffusion constant by an accurate
analytical study. First of all, the ring-collision effect is taken into full
consideration to calculate the hydrodynamics contribution by means of the
linearizing hydrodynamics approach. We then combine the kinetics
contribution to get two coupled equations for both the VACF and the
diffusion constant and to get their explicit solutions. Finally, the
obtained analytic results are carefully scrutinized with large scale
molecular dynamics simulations of the two-dimensional hard-disk fluid.

The key point of our consideration is that the kinetics and hydrodynamics
processes take place simultaneously. Without loss of generality, we suppose
that initially a tagged particle resides at the origin and moves along the $%
x $ axis with the momentum $p_{x}^{c}=mu_{x}^{c}$, where $u_{x}^{c}$ is its
initial velocity. $p_{x}^{c}$ also represents the memory of the particle's
initial moving direction. As the system evolves, upon the collisions
between the tagged particle and other particles, $p_{x}^{c}$ will transfer
to the latter. We refer to this process, i.e., the initial momentum of the
tagged particle transferring away from it, as the kinetic process, and
denote the portion of $p_{x}^{c}$ that has not transferred at time $t$ as $%
p_{x}^{k}(t)$. On the other hand, it is possible for the portion of $%
p_{x}^{c}$ that has transferred away to transfer back to the tagged particle
via ring collisions~\cite{Alder2}. Such a feedback is a collective effect of
the surrounding particles which we refer to as the hydrodynamic process and
denote the portion of $p_{x}^{c}$ that returns back to the tagged particle
at time $t$ as $p_{x}^{h}(t)$. Following the definition of the VACF, $%
C(t)=\langle u_{x}(t)u_{x}^{c}\rangle =\langle p_{x}(t)p_{x}^{c}\rangle
/m^{2}$, the VACF as well as the diffusion constant can be divided into two
parts respectively, i.e., $C(t)=C_{k}(t)+C_{h}(t)$ and $%
D(t)=D_{k}(t)+D_{h}(t)$.

We assume that $C_{k}(t)$ follows the kinetics prediction, i.e., $C_{k}(t)={%
\Omega }\exp ({-\Omega t/D_{k}})$~\cite{Dorfman2, Dorfman3}. To obtain $%
C_{h}(t)$ and $D_{h}(t)$, we calculate $p_{x}^{h}(t)$ in the following. Let $%
p(\mathbf{r},t)$ represent the density of the initial momentum transferring
to the unit volume at position $\mathbf{r}$ and time $t$ and the probability
with which the tagged particle appears in this area is $\rho (\mathbf{r},t)$%
; then on average the portion of the initial momentum returns to the tagged
particle is $\rho (\mathbf{r},t)p(\mathbf{r},t)/n$. The total amount of the
initial momentum returns to it is thus
\begin{equation}
p_{x}^{h}(t)=(1/n)\int p(\mathbf{r},t)\rho (\mathbf{r},t)d\mathbf{r}.
\label{2}
\end{equation}%
In this way, the ring-collision effect of various orders is integrated
exhaustively.

We calculate $p(\mathbf{r},t)$ by hydrodynamics approaches. With the
conventional assumption that the mean velocity of the fluid is zero and the
local deviations of a hydrodynamic variable from its average value are
small, we get $\mathbf{p}(\mathbf{r},t)=mn\mathbf{u}(\mathbf{r},t)=m\mathbf{j%
}(\mathbf{r},t)$, where $\mathbf{j}(\mathbf{r},t)$ is the local particle
current. The relaxation properties of the momentum is thus connected to the
particle current. We then have the conservation equations of the total
particle number, the energy, and hte momentum of the system under this
assumption, and by linearizing them with the help of the double transforms
with respect to space (Fourier) and time (Laplace), we get the hydrodynamics
equations (see, e.g., Ref.~\cite{theorySP}). This is conventional in the
hydrodynamics approaches, but our particular treatment~\cite{SM} is to solve
the equations with the specific initial conditions of
$[p_{x}(0,0),p_{y}(0,0),p_{z}(0,0)]$ $=[p_{x}^{c}\delta (\mathbf{r}),0,0]$, $%
\Delta T(0,0)=\Delta T\delta (\mathbf{r})$, and\ $\Delta n(0,0)=\Delta
n\delta (\mathbf{r})$ for a three dimensional fluid, which gives the heat mode
\begin{equation}
\frac{p^{heat}(\mathbf{k},t)}{p_{x}^{c}}=\frac{k_{y}^{2}+k_{z}^{2}}{k^{2}}%
\exp (-\nu _{k}k^{2}t)  \label{3}
\end{equation}%
and the sound mode%
\begin{equation}
\frac{p^{sound}(\mathbf{k},t)}{p_{x}^{c}}=\frac{k_{x}^{2}}{k^{2}}\exp
(-\Gamma k^{2}t)\cos (c_{s}kt)  \label{4}
\end{equation}%
of the momentum (or particle current) fluctuation under the long wave
approximation (up to the $\sim k^{2}$ term). Here $c_{s}$ is the sound speed
and the parameter $\Gamma $ represents the sound attenuation coefficient
defined in Ref.~\cite{theorySP}, p.~229. We then get $p(\mathbf{k},t)=
p^{heat}(\mathbf{k},t)+p^{sound}(\mathbf{k},t)$. These solutions apply to
a two dimensional fluid as well with $k_{z}=0$.

As to $\rho (\mathbf{r},t)$, the kinetics theory gives $\rho (\mathbf{r},t)=%
\frac{1}{4\pi D_{k}t}\exp ({-\frac{r^{2}}{4D_{k}t}})$. Considering that the
hydrodynamics effect enhances the diffusion due to the returning of the
initial momentum, we assume that $\rho (\mathbf{r},t)$ generally has the
same form as given by the kinetics theory but with a time-dependent
diffusion constant $D_{h}(t)+D_{k}$ instead. Transforming it with respect to
the space, we get $\rho (\mathbf{k},t)=\exp [-(D(t)+D_{k})k^{2}t]$.

To find the solution of Eq.~(2) we employ the Parseval's formula $%
\int_{-\infty }^{\infty }p(\mathbf{r},t)\rho (\mathbf{r},t)d\mathbf{r}=(%
\frac{1}{2\pi })^{2}\int_{-\infty }^{\infty }p(\mathbf{k},t)\rho (\mathbf{k}%
,t)d\mathbf{k}$ (see e.g., Ref~\cite{Rudin}, p.~187) to convert the intergal
in the configuration space to that in the wave-vector space, which gives
\begin{equation}
C_{h}(t)=\frac{\Omega(d-1)}{nd}[4\pi (D_{h}(t)+D_{k}+\nu _{k})t]^{-\frac{d%
}{2}}.  \label{5}
\end{equation}%
Note that in deriving this result the sound mode is
neglected as its contribution is a high order small quantity~\cite{SM}. Eq.~(5)
will be the same as that given by the conventional approaches~\cite{theorySP,
Dorfman1, Dorfman2} if one sets $D_{h}(t)=0$, but there it is an approximation
result under the cut-off assumption (see e.g., Ref.~\cite{theorySP}, p. 248).

In deriving Eq.~(5), the underlying assumption is that the initial momentum $%
p_{x}^{c}$ of the tagged particle has transferred to the surrounding
particles completely. At short times, the portion of $p_{x}^{c}$ that the
hydrodynamics process accounts for is $p_{x}^{c}[1-\exp (-{\Omega t/}D^{k})]$%
; the remaining portion is still carried by the tagged particle. For this
reason Eq.~(5) can be extended straightforwardly to the short time regime as
\begin{equation}
C_{h}(t)=\frac{\Omega(d-1)}{nd}(1-e^{-\frac{{\Omega }t}{D_{k}}})[4\pi
(D_{h}(t)+D_{k}+\nu_{k})t]^{-\frac{d}{2}}.  \label{6}
\end{equation}
This extended formula therefore applies to the whole time range. Inserting $%
C_{h}(t)$ into the Green-Kubo formula and solving the coupled equations, we
get
\begin{eqnarray}
C_{h}(t) &=&\frac{\Omega(d-1)}{nd}(1-e^{-\frac{{\Omega }t}{D_{k}}})(4\pi t)^{-%
\frac{d}{2}}A(t)^{-\frac{d}{d+2}}  \label{7} \\
D_{h}(t) &=&A(t)^{\frac{2}{d+2}}-(D_{k}+\nu_{k})  \label{8}
\end{eqnarray}%
%
%
%
with $A(t)=(D_{k}+v_{k})^{2}+[G+\log (t/D_{k})-ei(-{\Omega } t/D_{k})]{\Omega }
/(4\pi n)$ for $d=2$ and $A(t)=(D_{k}+\nu _{k})^\frac{5}{2}+(5/12{\ \pi }%
n)D_{k}^{-1/2}\Omega^{2/3} erf [({\Omega t}/D_{k})^{1/2}]-(5/12n) \Omega \pi
^{-3/2} [1-\exp (-\Omega t/D_{k})]t^{-1/2}$
for $d=3$. Here $G=0.577...$ is the Euler-Gamma constant, $erf$ and $ei$
are respectively the error function and the exponential integral function.

\begin{figure}[tbp]
\centering
\includegraphics[width=8.6cm]{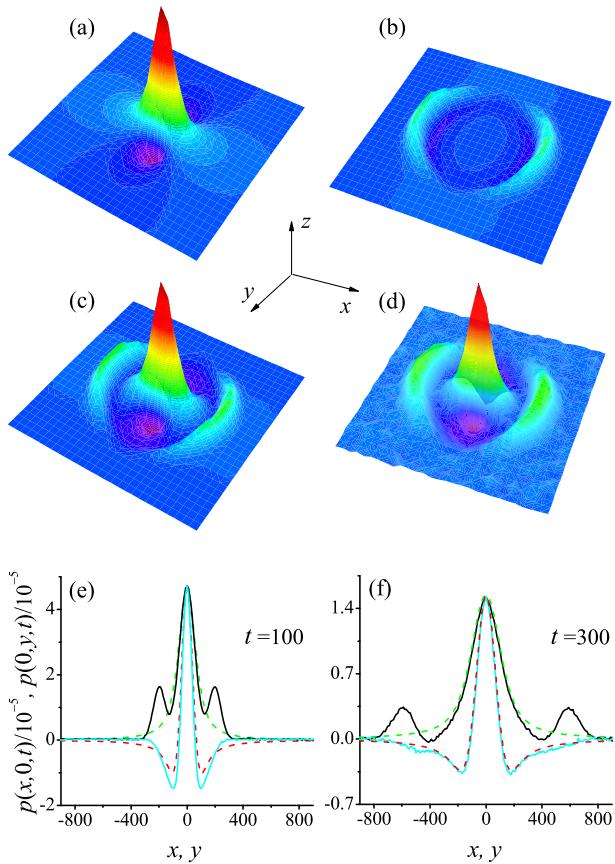} \vskip-0.5cm
\caption{The analytical result of the heat mode (a), the sound mode (b), and
their superposition (c) at $t=300$ for $\protect\nu _{k}=7.7$. (d) The
simulated $p(\mathbf{r},t)$ at $t=300$; (e)-(f) The intersections of the
simulated $p(\mathbf{r},t)$ with $x=0$ (black solid lines) and $y=0$ (cyan
solid lines) at $t=100$ and $300$, respectively. They are compared with the
heat mode given by Eq.~(9) with the best fitting value of $\protect\nu %
_{k}=8.3$ whose intersections with $x=0$ ($y=0$) are indicated by the green
dashed (red dashed) lines. For simulations $\protect\sigma =4$.}
\end{figure}

For $d=2$, the hydrodynamics contribution of $D_{h}(t)$ diverges in the
thermodynamical limit. The asymptotic solutions of Eq.~(7)-(8) are $C(t)={%
\Omega }\sqrt{1/16\pi n}(t\sqrt{\ln t})^{-1}$ and $D(t)=\sqrt{\ln t/(4\pi
n\Omega )}$, which agree with previous results~\cite{wain-tlnt, tlnt}. From
Eq. (6) one can further deduce that these solutions apply beyond the time
scale of $t>t_{self}=\exp [(4\pi n/\Omega )(D_{k}+\nu _{k})^{2}]$ when $%
D_{k}+\nu _{k}$ can be ignored comparing to $D_{h}(t)$. For $t<t_{self}$,
the VACF experiences a transition from $\sim t^{-1}$ to $\sim (t\sqrt{\ln t}%
)^{-1}$. Particularly, in the period that $D_{h}(t)$ is negligible
comparing to $D_{k}+\nu _{k}$, the tail of the VACF can be well approximated
with $\sim t^{-1}$. For $d=3$, the VACF converges asymptotically to $\sim
t^{-3/2}$, and $D_{h}(t)$ converges to the constant ${D_{h}(\infty )=[{%
\left( D_{k}+\nu _{k}\right) }^{5/2}+\frac{5}{12n{\pi }}(D_{k})^{-1/2}\Omega
^{3/2}]}^{2/5}-(D_{k}+\nu _{k})$ which gives the upper bound of the
hydrodynamics contribution.

In the following we put these analytical results into numerical tests with
the hard-disk fluid model. This model is the simplest paradigmatic fluid
model but has general importance for fluids since its structure do not
differ in any significant way from that corresponding to more complicated
interatomic potentials~\cite{theorySP}. It consists of $N$ disks of unitary
mass $m=1$ moving in an $L\times L$ square box with the periodic boundary
conditions. As the VACF is free from the finite-size effects for time $%
t<t_{f}=L/(2c_{s})$~\cite{pre2008, chenfinite, SM}, the box should be big
enough to guarantee the computed tail of the VACF to be accurate. In our
simulations the system size is fixed at $L=2000$ and $N=40000$ throughout.
As such the average disk number density is fixing at $n=0.01$ and the disk
diameter, $\sigma$, is adopted to control the packing density $\phi =n\pi
\sigma ^{2}/4$ (referred to as the density for short in the following). In
particular, $\sigma=2$ to $9$ is investigated numerically that covers both
the regimes of gas and liquid. As a reference, the crystallization density
is $\phi=0.71$, corresponding to $\sigma =9.5$. The system is evolved with
the event-driven algorithm~\cite{Alder, cal} at the dimensionless
temperature $T=1$ ($k_{B}$ is set to be unity).

\begin{table}[tb]
\begin{tabular}{cccccc}
\tableline $\sigma (\phi )$ & 2(0.03) & 4(0.13) & 6(0.28) & 8(0.50) & 9(0.63)
\\
\tableline $\nu _{k}$(E) & 14.1 & 7.7 & 6.3 & 8.5 & 14.3 \\
$\nu _{k}$(S) & 14.3 & 8.3 & 8.05 & 15.0 & 35.0 \\
$c_{s}$(E) & 1.51 & 1.85 & 2.72 & 5.50 & 9.97 \\
$c_{s}$(S) & 1.65 & 1.97. & 2.81 & 5.71 & 9.85 \\
$D_{k}$(E) & 13.4 & 5.70 & 2.76 & 1.10 & 0.59 \\
$D_{k}$(S) & 13.61 & 5.50 & 2.35 & 0.54 & 0.27 \\
$D^{h}(t)/D^{k}$ & 0.14 & 0.69 & 2.09 & 7.29 & 7.32 \\
\tableline &  &  &  &  &
\end{tabular}%
\caption{Comparison of the kinetics coefficients for the hard-disk fluid 
model obtained by the Enskog formula (E) and by our analytical prediction
based on the simulation results of $p(\mathbf{r},t)$ (S). $t=10^7$ for 
$D^h (t)$ in the last row.}
\label{T1}
\end{table}

\begin{figure*}[tbp]
\centering
\includegraphics[width=18cm]{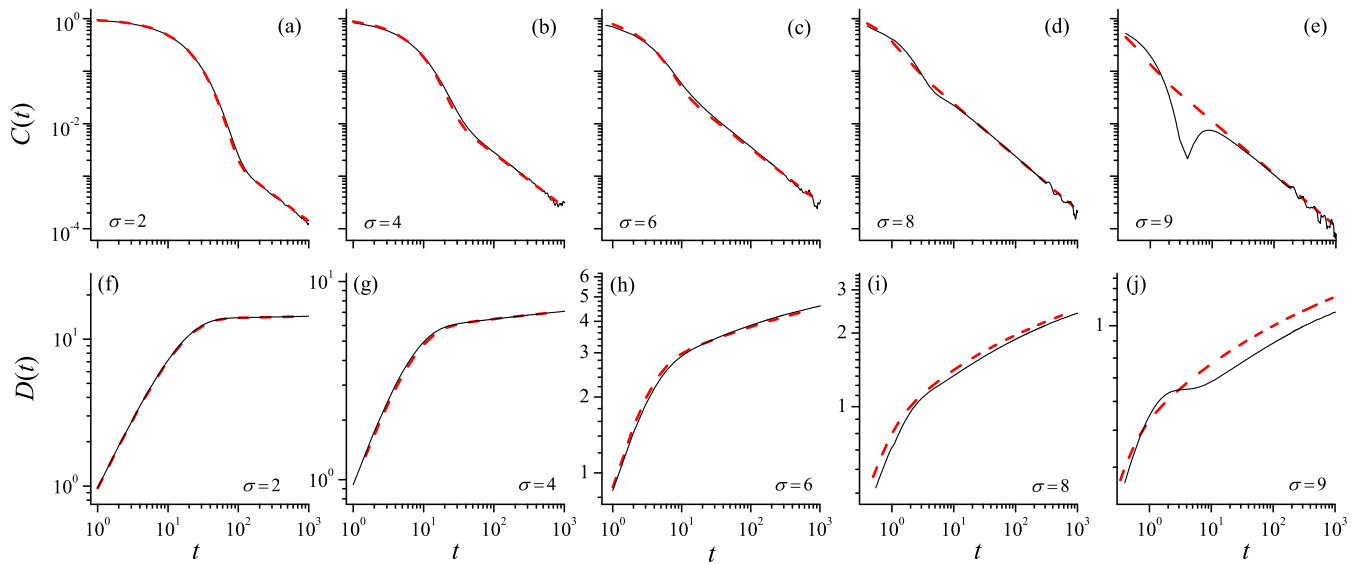} \vskip-0.3cm
\caption{(Color online) (a)-(e) Comparison of the simulated VACFs (black
solid lines) and the analytical results with the parameters $\protect\nu %
_{k} $ and $D_{k}$ given by our computations (red dotted lines) at various
system densities. (f)-(j) The same as (a)-(e) but for the diffusion constant.
}
\end{figure*}

We first check the prediction of the hydrodynamics modes. For the heat mode,
the inverse Fourier transform of Eq.~(3) with respect to space gives
\begin{equation}
\frac{p^{heat}(\mathbf{r},t)}{p_{x}^{c}}=\frac{x^{2}-y^{2}}{2\pi r^{4}}%
(1+e^{-\frac{r^{2}}{4\nu _{k}t}})+\frac{y^{2}}{4\pi r^{2}\nu _{k}t}e^{-\frac{%
r^{2}}{4\nu _{k}t}}.  \label{9}
\end{equation}%
Similarly, the sound mode can be obtained by the inverse transform of
Eq.~(4) (in form of series expansion~\cite{SM}). Combining them together, we
then have the analytical result of $p(\mathbf{r},t)$. An example for $\nu
_{k}=7.7$ (the Enskog value for $\sigma =4$) is given in Fig.~1(a)-(c).
Noting that $p(\mathbf{r},t)$ is not isotopic. Numerically, it is computed
by the spatiotemporal correlation function $\langle \widetilde{p}(\mathbf{r}%
,t)p_{x}^{c}\rangle $ ~\cite{chendiffusion}:
\begin{equation}
\frac{p(\mathbf{r},t)}{p_{x}^{c}}=\frac{\langle \widetilde{p}(\mathbf{r}%
,t)p_{x}^{c}\rangle }{\langle |p_{x}^{c}|^{2}\rangle }+\frac{n}{N-1}.
\label{10}
\end{equation}%
Here $\widetilde{p}(\mathbf{r},t)$ represents the temporal momentem density.
See Fig.~1(d) for the simulated $p(\mathbf{r},t)$ for $\sigma =4$ as an
example.

The explicit expression of the heat mode allow us to compute the viscosity
diffusivity $\nu _{k}$ numerically based on our equilibrium simulations of $%
p(\mathbf{r},t)$. This can be done conveniently by best fitting the heat
mode, i.e, the center peak, of the simulated $p(\mathbf{r},t)$ [see
Fig.~1(d)] with $p^{heat}(\mathbf{r},t)$. In this way, we find that the $\nu
_{k}$ value for the illustrating case of $\sigma =4$ is $\nu _{k}=8.3$. Note
that this `measurement' does not depend on time [see Fig.~1(e)-(f) for two
different times], implying that the viscosity diffusivity is not affected by
the hydrodynamics effect. In fact, previous numerical studies using the
Helfand-Einstein formula have shown that it does not depend on the system
size either~\cite{formula-2d, 2d-tailnotimport}, which also supports that
the viscosity diffusivity is a time-independent constant. Table~I summarizes
the value of $\nu _{k}$ computed in this way for various system densities
and the sound speed computed by tracing the sound mode (i.e., side peaks) of
the simulated $p(\mathbf{r},t)$. It is important to notice that the $\nu _{k}$
value we obtain agrees very well with that given by the Enskog formula (under
the first Sonine polynomial approximation~\cite{formula-2d, Gass, pl-eskog})
at the dilute gas regime but deviates remarkably as the density increases. In
contrast, the sound speed values agrees with each other very well for all the
densities.

To compare the simulated $p(\mathbf{r},t)$ with the analytical prediction, we
assume the numerically measured value of $\nu _{k}$ in the latter and find the
agreement is perfect in both the gas and liquid regimes even in a very dense
liquid case ($\sigma =8$). But as expected, if the Enskog value of $\nu _{k}$
is taken, then the agreement is perfect only in the dilute gas case. For a
moderate density the agreement can be good qualitatively with noticeable
difference [compare Fig.~1(c) and (d)].

Next we turn to the instantaneous diffusion constant. Numerically it can be
computed by the Green-Kubo formula, $D(t)=\int_{0}^{t}C(t^{\prime
})dt^{\prime }$, if $C(t)$ is calculated. It can also be calculated
according to its definition, $D(t)=d\langle r^{2}(t)\rangle /dt$, by tracing
the tagged particle directly for $\langle r^{2}(t)\rangle $. The two ways
give the same result~\cite{SM}. Figure~2(a)-(e) show the VACF at different
densities, whose tail is close to $\sim t^{-1}$ but deviates differently. 
Taking the Enskog kinetics coefficients given in Table I, we have $t_{self}=$ 
$10^{23}$, $10^{10}$, $10^{4}$, $10^{13}$, and $10^{35}$ for the corresponding 
densities. The value of $t_{self}$ is big at the dilute and the dense limit 
because in the former $D_{k}$ is big and in the latter $\nu _{k}$ becomes big. 
In both cases $D_{h}(t)$ is comparatively small in a remarkably long time 
range in which it can be neglected and $C(t)\sim t^{-1}$ is expected 
[see Eq.~(6)]


We have evidenced the considerable deviation of the computed $\nu _{k}$ from
the Enskog value. For $D_{k}$ this is also the case. To evaluate $D_{k}$, 
we have $D(t)=D_h(t)+D_k=A(t)^\frac{2}{d+2}-\nu_k$ [see Eq.~(6)], hence by
taking the simulated $D(t)$ and numerically measured $\nu_k$  we can solve 
$A(t)$ and thus $D_k$. In doing so we have to adopt a big enough value of $t$ 
to make sure that the solved $D_k$ is a constant whose value does not change
if $t$ is increased further. The $D_k$ value obtained in this way is presented
in Table I; it can be seen that it is close to the corresponding Enskog value 
at low densities but deviates increasingly again as the density increases.

Applying the values of $D_{k}$ and $\nu _{k}$ measured in our method, we find
that the analytical results of $C(t)$ and $D(t)$ agree perfectly with simulated
ones, except an obvious disagreement in the VCAF at the highest density we
have investigated ($\sigma =9$): a dip appears around the transition point $%
t\approx \tau $ and as a result the predicted $D(t)$ is slightly larger than
the simulated result for $t>\tau $ [Fig.~2(j)]. The dip may be induced
by the lattice feature as the system is close to the crystallization phase.
Nevertheless, the diffusion constant can still be predicted fairly well due
to the long power law tail that becomes dominant.

It is useful for application aims to have an estimation of the
hydrodynamics effect in a macroscopic system. The average distance between
two neighboring molecules in the air is about $10^{-9}$ meter; given that
our model has a macroscopic size, say one centimeter, we have $L\sim 10^{8}$
and as such the time a particle diffuses freely without being influenced by
the boundaries is $t_{f}\sim 10^{7}$. In Tab.~I the ratio $%
D_{h}(10^{7})/D_{k}$ for the hard-disk fluid is listed, from which it can be
seen that in a dilute gas the kinetics contribution dominates, but as the
density increases, the hydrodynamics contribution increases dramatically and
the kinetics contribution turns out to be negligible.

The hydrodynamics influence is much weaker for a three-dimensional system
because of the fast convergence of the Green-Kubo integral. 
Adopting the kinetics coefficients of the hard-disk fluid for an
estimation, we find ${D_{h}(\infty )/D^{k}<}1\%$ in general. Only in certain
extreme situations, for example $D^{k}\rightarrow 0$ in the crystallization
limit, the hydrodynamics contribution may become comparable to that of the
kinetics.

In summary, based on the consideration that the kinetics and hydrodynamics
processes take place simultaneously and by characterizing them with the
losing and returning of the memory to the initial state of a particle, we
have derived explicitly the VACF and the diffusion constant of a simple
fluid, which are firmly corroborated by the numerical study of the hard-disk
fluid model. It is found that (1) in the two-dimensional case, the
hydrodynamics influence to the particle diffusion is negligible in the
dilute gas regime but becomes dominant at high densities. For a
three-dimensional fluid, the hydrodynamics influence is in general
negligible; (2) The relaxation of momentum is not isotropic. This is
different from the relaxation of mass-density fluctuations~\cite{theorySP}.
(3) Our simulations show that the Enskog formula should be improved for
calculating kinetics coefficients when the system density is high.

This work is supported by the National Natural Science Foundation of China
(Grant No. 11335006) and the NSCC-I computer system of China.

\end{document}